# An integrative sparse boosting analysis of cancer genomic commonality and difference

**Yifan Sun,[1] Zhengyang Sun,[1] Yu Jiang,[2] Yang Li[1] and Shuangge Ma[1,3]**

## Abstract

In cancer research, high-throughput profiling has been extensively conducted. In recent studies, the integrative analysis of data on multiple cancer patient groups/subgroups has been conducted. Such analysis has the potential to reveal the genomic commonality as well as difference across groups/subgroups. However, in the existing literature, methods with a special attention to the genomic commonality and difference are very limited. In this study, a novel estimation and marker selection method based on the sparse boosting technique is developed to address the commonality/difference problem. In terms of technical innovation, a new penalty and computation of increments are introduced. The proposed method can also effectively accommodate the grouping structure of covariates. Simulation shows that it can outperform direct competitors under a wide spectrum of settings. The analysis of two TCGA (The Cancer Genome Atlas) datasets is conducted, showing that the proposed analysis can identify markers with important biological implications and have satisfactory prediction and stability.

## 1 Introduction

In cancer research, high-throughput profiling has been extensively conducted, generating a large amount of genomic data. A major goal of genomic studies is to identify markers associated with cancer outcomes/phenotypes. A vast literature has been published. In most of the existing studies especially the early ones, the focus has been on a homogenous set of cancer patients. With the accumulation of data and evidences, there has been a growing attention to the "relationships" across different/heterogeneous cancer patient groups.[1,2] Here different groups may have different cancer types, subtypes, or biomarker values. As a representative example, building on The Cancer Genome Atlas (TCGA) Research Network, the National Cancer Institute (NCI) initiated the Pan-cancer project to examine the commonality and differences across cancer types.[3] Other examples are presented in literature.[4–7] There are two equally important aspects of such studies. The first aspect is difference. Cancer is highly heterogeneous. Different cancer types/subtypes are expected to have different genomic basis. Even patients with the same cancer type/subtype often have diverse genomic markers, which can be reflected in differences in biomarker values. The second aspect is commonality: some genetic mutations, epigenetic changes, pathways, and others have been identified as playing important roles in multiple cancers. Most cancers share similar fundamental properties such as metastasis. To comprehensively understand cancer, it is important to identify both genomic difference and commonality.

With heterogeneous cancer patient groups, the most straightforward approach is to conduct meta-analysis, under which each group is analyzed separately, and the results are pooled and compared.[8] Cancer genomic studies

[1]Center for Applied Statistics, School of Statistics, Renmin University of China, Beijing, China
[2]School of Public Health, University of Memphis, Tennessee, USA
[3]Department of Biostatistics, Yale School of Public Health, New Haven, CT, USA

**Corresponding author:**
Shuangge Ma, Department of Biostatistics, Yale School of Public Health, 60 College ST, New Haven 06520, CT , USA.
Email: shuangge.ma@yale.edu

usually have the "small sample size, high-dimensional measurement" characteristic, leading to unsatisfactory results for each individual group and hence overall meta-analysis.[9–11] In the recent studies, integrative analysis has been proposed as a viable solution and shown to outperform meta-analysis. Under integrative analysis, the raw data from multiple groups (datasets) are pooled prior to analysis. The heterogeneity across groups is taken into account in model estimation and marker selection.[12] In "standard" integrative analysis, there is a lack of attention to the relationships across groups/datasets. There are also a few integrative analysis approaches that pay special attention to the similarity across groups/datasets. For example, the contrasted penalization approach[13] applies a penalty to smooth the regression coefficients of the same covariates across multiple datasets. Penalization[14] and sparse boosting[15] techniques have been adopted to promote similarity in model sparsity structures. It is noted that the aforementioned and some other approaches can achieve similarity but not commonality. That is, the identified effects/models may have similar but not exactly identical magnitudes. In the literature, the most relevant is the analysis conducted in Sun et al.[16] which also considers commonality and difference and adopts the penalization technique.

In this study, data on heterogeneous cancer patient groups are taken in consideration. The goal, beyond the "standard" estimation and variable selection, is to identify the genomics commonality and difference across groups. Compared to the single-group analysis, the proposed analysis can be more informative by revealing the common mechanisms as well as group-specific cancer genomic characteristics. Advancing from meta-analysis and integrative analysis (including those techniques that promote similarity), the proposed analysis can directly obtain commonality, that is, the shared common effects can have exactly equal, as opposed to similar, estimates. In both high-dimensional and low-dimensional statistical learning, it has been observed that there is no dominatingly better technique. The proposed approach is based on sparse boosting, which is one of the most recent and highly effective boosting techniques.[17] Published studies[18,19] have shown that sparse boosting has certain unique advantages (and also possible disadvantages) compared to penalization and other high-dimensional techniques. It is thus of interest to develop the proposed method beyond the penalization one.[16]

The rest of the article is organized as follows. In Section 2, we describe the data and model settings. The proposed sparse boosting approach is described in Section 3. In Section 4, we conduct simulation and comparison with the alternatives. The proposed approach is also applied to two TCGA datasets. Section 5 concludes this paper. Additional technical details and numerical results are provided in Appendix.

## 2 Data and model settings

Consider the integrative analysis of $M$ independent datasets. Each dataset may correspond to a different type/subtype of cancer or a different subpopulation of the same type of cancer. Let $\boldsymbol{Y}^1, \boldsymbol{Y}^2, \ldots, \boldsymbol{Y}^M$ be the response variables and $\boldsymbol{X}^1, \boldsymbol{X}^2, \ldots, \boldsymbol{X}^M$ be the genomic measurements. In our numerical study, we analyze gene expression data. It is noted that the proposed approach can also be applied to other types of omics measurements. To simplify notation, assume that $M$ datasets measure the same set of covariates. With minor modifications, the proposed approach can also accommodate partially matched covariate sets. In dataset $m(=1, 2, \ldots, M)$ with $n^m$ i.i.d. observations, $\boldsymbol{Y}^m$ is associated with $\boldsymbol{X}^m$ via $\boldsymbol{Y}^m \sim \Phi(\boldsymbol{X}^m \boldsymbol{\beta}^m)$, where $\boldsymbol{\beta}^m = (\beta_1^m, \beta_2^m, \ldots, \beta_p^m)^\top$ is the $p$-vector of regression coefficients. Model $\Phi$ is assumed to have a known form. In numerical study, we consider continuous outcomes under linear regression (LR) models and right censored survival outcomes under accelerated failure time (AFT) models. Details on the model settings and estimation objective functions are provided in Appendix A. It is noted that, with minor modifications, the proposed approach can also accommodate other types of outcomes/models. We consider the setting with $p > n^m$, $m = 1, 2, \ldots, M$, which matches practical cancer genomic data.

A common limitation shared by many of the existing studies is that there is insufficient attention to the "interconnections" among genomic measurements. Specifically, genes form functional groups, with those in the same groups tending to function coordinately and those in different groups behaving differently. In our analysis, the $p$ genomic measurements are assumed to belong to $K$ non-overlapping groups, with $p_{(k)}$ in group $k$. Here the groups can be constructed functionally (for example, based on pathway or GO information) or statistically (for example, via clustering). Let $\boldsymbol{\beta}_{(k)}^m$ denote the coefficient vector of covariates in group $k$ in dataset $m$. If $\boldsymbol{\beta}_{(k)}^1 = \boldsymbol{\beta}_{(k)}^2 = \cdots = \boldsymbol{\beta}_{(k)}^M$, then group $k$ represents a commonality. In contrast, if there are datasets $m_1$ and $m_2$ with $\boldsymbol{\beta}_{(k)}^{m_1} \neq \boldsymbol{\beta}_{(k)}^{m_2}$, then group $k$ represents a difference. It is noted that when $M > 2$, it is possible to define partial commonality (difference). Details are provided below. Our goal is to properly discriminate commonality and difference in estimation and variable selection.

## 3 Methods

The proposed approach is based on the boosting technique. Boosting is a generic statistical learning technique that aggregates a series of weak learners into a strong one. A long array of published studies show that boosting has many desirable advantages, including flexibility, implementation simplicity, affordable computational cost, and satisfactory prediction performance.[20,21] Boosting can accommodate multiple types of weak learners. In this article, we consider regression models with linear covariate effects and linear functions of covariates as weak learners, to achieve simple interpretability.

### 3.1 Sparse boosting

In the analysis of cancer genomic data, the selection of relevant genomic measurements is as important as model building. With the ordinary boosting, variable selection can be achieved with early stopping. However, it has been suggested that the resulted models are not "sparse enough". Sparse boosting (SBoost) has been designed to further achieve sparsity.[17] It imposes penalty to promote the selection of sparser models. As SBoost is the technical basis of the proposed approach, it is presented below for the completeness of this article.

Consider the analysis of the $m$th dataset. Use $L^m(\cdot)$ to denote the loss function. For the LR and AFT models, its form is described in Appendix A. The SBoost algorithm proceeds as described in Algorithm 1.

---

**Algorithm 1** SBoost

---

**Step 1**: Initialization. Denote $\boldsymbol{\beta}^{m[t]}$ as the estimate of $\boldsymbol{\beta}^m$ in the $t$th iteration. Set $t=0$. Initialize $\boldsymbol{\beta}^{m[t]}=0$.
**Step 2**: Fit and update. $t=t+1$.
Compute $(\hat{s},\hat{\gamma})=\arg\min_{1\le s\le p,\gamma}\left\{L^m(\boldsymbol{\beta}^{m[t-1]}+\gamma\boldsymbol{e}_s)+\frac{\log n^m}{n^m}\sum_{j=1}^{p}\text{II}(\beta_j^{m[t-1]}+\gamma e_{s,j}\neq 0)\right\}$, where $\boldsymbol{e}_s$ is a $p$-vector with the $s$th component equal to 1 and others equal to 0, and $e_{s,j}$ is its $j$th component.
Update $\boldsymbol{\beta}^{m[t]}=\boldsymbol{\beta}^{m[t-1]}+\nu\hat{\gamma}\boldsymbol{e}_{\hat{s}}$.
**Step 3**: Iteration. Repeat Step 2 for $T$ times.
**Step 4**: Stopping. At the $t$th iteration, compute $F^{m[t]}=L^m(\boldsymbol{\beta}^{m[t]})+\frac{\log n^m}{n^m}\sum_{j=1}^{p}\text{II}(\beta_j^{m[t]}\neq 0)$. Select the optimal number of iterations as $\hat{t}=\arg\min_{1\le t\le T}F^{m[t]}$.
**Step 5**: Output. The regression coefficients are estimated as $\boldsymbol{\beta}^{m[\hat{t}]}$. The strong learner for dataset $m$ is $f^m=\boldsymbol{X}^m\boldsymbol{\beta}^{m[\hat{t}]}$.

---

The key difference between SBoost and ordinary boosting is that the SBoost's objective function has two components. The first is the lack-of-fit measure, as in ordinary boosting. The second component explicitly penalizes model complexity and hence can promote sparsity. In the literature, multiple model complexity measures have been used in SBoost, including BIC, AIC, MDL (minimum description length), and others. In Algorithm 1, we use a BIC type measure as an example, which has been adopted in multiple published studies. $\nu$ is the step size. Published studies suggest that its value is not crucial as long as it is not too big. In our numerical analysis, $\nu$ is set to be 0.1 following Buhlmann and Yu.[17]

### 3.2 Sparse boosting for multiple datasets

Consider the analysis of $M$ independent datasets using the SBoost technique. The integrative sparse boosting (Int-SBoost) algorithm proceeds as described in Algorithm 2.

---

**Algorithm 2** Int-SBoost

---

**Step 1**: Initialization. Set $t=0$. Initialize $\boldsymbol{\beta}^{m[t]}=0,\ m=1,2,\ldots,M$.
**Step 2**: Fit and update. $t=t+1$. For $m=1,2,\ldots,M$, compute

$$(\hat{s},\hat{\gamma})=\underset{1\le s\le p,\gamma}{\arg\min}\left\{L^m(\boldsymbol{\beta}^{m[t-1]}+\gamma\boldsymbol{e}_s)+\frac{\log n^m}{n^m}\sum_{j=1}^{p}\text{II}(\beta_j^{m[t-1]}+\gamma e_{s,j}\neq 0)\right\}$$

Update $\boldsymbol{\beta}^{m[t]} = \boldsymbol{\beta}^{m[t-1]} + \nu\hat{\gamma}^m \boldsymbol{e}_{\hat{s}}$, where $\nu$ is the step size as in SBoost.
**Step 3**: Iteration. Repeat Step 2 for $T$ times.
**Step 4**: Stopping. At the $t$th iteration, compute $F^{[t]} = \sum_{m=1}^{M}\left\{L^m(\boldsymbol{\beta}^{m[t]}) + \frac{\log n^m}{n^m}\sum_{j=1}^{p} \mathrm{II}(\boldsymbol{\beta}_j^{m[t]} \neq 0)\right\}$. Select the optimal number of iterations as $\hat{t} = \arg\min_{1\leq t\leq T} F^{[t]}$.
**Step 5**: Output. The regression coefficients are estimated as $\boldsymbol{\beta}^{m[\hat{t}]}$. The strong learner for dataset $m$ is $f^m = \boldsymbol{X}^m\boldsymbol{\beta}^{m[\hat{t}]}$.

As an integrative analysis method, Int-SBoost simultaneously takes multiple datasets into consideration. In each iteration, Int-SBoost applies sparse boosting to each dataset separately. It differs from SBoost in that, in deciding the stopping rule, all datasets are considered together. Specifically, all datasets have the same selected number of iterations, roughly corresponding to the same amount of regularization. It is noted that it is possible to design an Int-SBoost method to also take multiple datasets into consideration in Step 2.

## 3.3 Sparse boosting to identify commonality and difference across datasets

Although Int-SBoost can conduct the integrative analysis of multiple datasets, it does not have an explicit mechanism to identify commonality and difference across datasets. To this end, we propose the CD (commonality and difference)-SBoost algorithm as described in Algorithm 3.

**Algorithm 3** CD-SBoost

**Step 1**: Initialization. Set $t=0$. Denote $\boldsymbol{\beta} = (\boldsymbol{\beta}^1, \boldsymbol{\beta}^2, \ldots, \boldsymbol{\beta}^M)$, and $\boldsymbol{\beta}^{[t]}$ as the estimate of $\boldsymbol{\beta}$ in the $t$th iteration. Initialize $\boldsymbol{\beta}^{[t]} = 0$.
**Step 2**: Fit and update. $t = t + 1$.
(I). For $s = 1, 2, \ldots, p$, determine the candidate set $\Gamma_s$ of $\gamma = (\gamma^1, \gamma^2, \ldots, \gamma^M)$. Denote $k_s$ as the group that covariate $s$ belongs to.
Case (a): For any $m_1 \neq m_2$, $\boldsymbol{\beta}_{(k_s)}^{m_1[t-1]} \neq \boldsymbol{\beta}_{(k_s)}^{m_2[t-1]}$. Compute $\tilde{\gamma}^m = \arg\min_\gamma L^m(\boldsymbol{\beta}^{m[t-1]} + \gamma\boldsymbol{e}_s)$. Then $\Gamma_s = \{(0, \ldots, 0, \tilde{\gamma}^m, 0, \ldots, 0)|m = 1, 2, \ldots, M\}$.
Case (b): There are $m_1, m_2, \ldots, m_l$ ($l \geq 2$), such that $\boldsymbol{\beta}_{(\boldsymbol{k}_s)}{}^{m_1[t-1]} = \boldsymbol{\beta}_{(\boldsymbol{k}_s)}{}^{m_2[t-1]} = \cdots = \boldsymbol{\beta}_{(\boldsymbol{k}_s)}{}^{m_l[t-1]}$. Compute $\tilde{\gamma}^A = \arg\min_\gamma \sum_{m\in A} L^m(\boldsymbol{\beta}^{m[t-1]} + \gamma\boldsymbol{e}_s)$, where $A$ is a non-empty subset of $\{m_1, \ldots, m_l\}$. Define $\mathcal{A} = \{A|A \subset \{m_1, \ldots, m_l\}\} \setminus \emptyset$. Then $\Gamma_s = \{\tilde{\gamma}^A \boldsymbol{I}_A | A \subset \mathcal{A}\}$, where $\boldsymbol{I}_A$ is a length-$M$ vector with the components in set $A$ equal to 1 and others equal to 0.
(II). Compute

$$(\hat{s}, \hat{\gamma}) = \underset{1\leq s\leq p, \gamma\in\Gamma_s}{\arg\min}\left\{\sum_{m=1}^{M}\left[L^m(\boldsymbol{\beta}^{m[t-1]} + \gamma^m\boldsymbol{e}_s) + \frac{\log n^m}{n^m}\sum_{j=1}^{p}\mathrm{II}\,(\boldsymbol{\beta}_j^{m[t-1]} + \gamma\boldsymbol{e}_{s,j} \neq 0)\right] + pen\left(\boldsymbol{\beta}^{[t-1]} + \sum_{m=1}^{K}\gamma^m\boldsymbol{E}_{s,m}; \lambda\right)\right\}$$

Here $pen(\boldsymbol{\beta}; \lambda) = \lambda \times \frac{\sum_{m_1\neq m_2}\sum_{k=1}^{K}\mathrm{II}(\boldsymbol{\beta}_{(k)}^{m_1} = \boldsymbol{\beta}_{(k)}^{m_2})}{\binom{M}{2}K}$, where $\lambda$ is the tuning parameter. $\boldsymbol{E}_{s,m}$ is a $p \times M$ matrix with the $(s,\, m)$th element equals to 1 and others equal to 0.

Update $\boldsymbol{\beta}^{[t]} = \boldsymbol{\beta}^{[t-1]} + \nu\sum_{m=1}^{M}\hat{\gamma}^m\boldsymbol{E}_{\hat{s},m}$, where the step size $\nu$ is set to be 0.1.
**Step 3**: Iteration. Repeat Step 2 for $T$ times.
**Step 4**: Stopping. At the $t$th iteration, compute

$$F(t) = \sum_{m=1}^{M}\left\{L^m(\boldsymbol{\beta}^{m[t]}) + \frac{\log n^m}{n^m}\sum_{j=1}^{p}\mathrm{II}(\boldsymbol{\beta}_j^{m[t]} \neq 0)\right\} + pen(\boldsymbol{\beta}^{[t]}; \lambda)$$

Select the number of iterations as $\hat{t} = \mathrm{argmin}_{1\leq t\leq T} F(t)$.
**Step 5**: Output. The regression coefficients are estimated as $\boldsymbol{\beta}(\hat{t}) = (\beta^1(\hat{t}), \beta^2(\hat{t}), \ldots, \beta^M(\hat{t}))$. The strong learner for dataset $m$ is $f^m = \boldsymbol{X}^m\boldsymbol{\beta}^m(\hat{t}),\ m = 1, 2, \ldots, M$.

Compared to the existing boosting based methods, including those described above, CD-SBoost approach has two major distinctions. The first is that a new penalty is developed to explicitly quantify difference across multiple datasets at the group level. Specifically, in $\frac{\sum_{m_1 \neq m_2} \sum_{k=1}^{K} \mathrm{II}(\boldsymbol{\beta}_{(k)}^{m_1} = \boldsymbol{\beta}_{(k)}^{m_2})}{\binom{M}{2} K}$, the numerator first counts the number of groups with identical regression coefficients in two distinct datasets, and then sums over all pairs of distinct datasets. The denominator is a normalization constant and counts the maximal number of identical groups across the $M$ datasets. With tuning parameter $\lambda$, this penalty takes value in $[0, \lambda]$. It is minimized if all groups behave the same across $M$ datasets (i.e. $\boldsymbol{\beta}_{(k)}^{1} = \cdots = \boldsymbol{\beta}_{(k)}^{M},\ 1 \leq k \leq K$), and is maximized if all groups behave differently. As such, this penalty can directly promote equal regression coefficients, i.e. commonality. It is also flexible enough to allow difference. The proposed analysis conducts group-based analysis: a whole group will be concluded as behaving same or differently across multiple datasets. Moreover, as a special case, if the analyzed datasets have a natural order, for example if they correspond to different cancer stages, then the penalty can be revised as $\lambda \times \frac{\sum_{m=1}^{M-1} \sum_{k=1}^{K} \mathrm{II}(\boldsymbol{\beta}_{(k)}^{m} = \boldsymbol{\beta}_{(k)}^{m+1})}{(M-1)K}$.

The second distinction is the computation of increments $\gamma = (\gamma^1, \gamma^2, \ldots, \gamma^M)$. Different from the existing integrative boosting approaches which determine the increments separately for each dataset, CD-SBoost determines all increments simultaneously. Specifically, for covariate $s$, a set $\Gamma_s$ is constructed which includes all candidate $\gamma$. If the regression coefficients of group $k_s$, which covariate $s$ belongs to, are identical in some datasets, say datasets $m_1, m_2, \ldots, m_l$ (that is, $\boldsymbol{\beta}_{(k_s)}^{m_1} = \ldots = \boldsymbol{\beta}_{(k_s)}^{m_l}$), some datasets in $\{m_1, m_2, \ldots, m_l\}$ are allowed to have identical increments in coefficients, so as to preserve commonality in these datasets. In contrast, if $\boldsymbol{\beta}_{(k_s)}^{m_1} \neq \boldsymbol{\beta}_{(k_s)}^{m_2}$ for any two distinct datasets $m_1$ and $m_2$, the increments $\gamma^m$'s are determined separately for each dataset $m$, and are allowed to differ for different $m$. For simplicity, we only update the regression coefficients of an individual dataset at a time.

*Tuning parameter selection*. For the selection of $\lambda$, we consider the high-dimensional BIC (HDBIC) criterion.[22] Specifically, consider $M$ independent datasets with sample sizes $n^1, \ldots, n^M$ and under linear regression models. Denote $\mathrm{df}^m$ as the number of important covariates in dataset $m$ and $\mathrm{df}^m$ as the residual sum of squares. The HDBIC is defined as

$$\mathrm{HDBIC} = \sum_{m=1}^{M} [n^m \log(\mathrm{RSS}^m / n^m) + \mathrm{df}^m \cdot \log(p) \log(n^m)]$$

It can be defined for other data and model settings accordingly. Compared to the ordinary BIC, HDBIC may generate sparser models and hence is more suitable for high-dimensional data.

## 3.4 A small example

To better appreciate the working characteristics of CD-SBoost, we consider a small simulation example. Specifically, three independent datasets are simulated, each with 50 subjects. $p = 200$ covariates belong to four non-overlapping groups. Among them, one group (group 1) behaves the same across datasets (that is, commonality), two groups (groups 3 and 4) behave "partially the same" (that is, they behave the same in two out of three datasets), and one group (group 2) behaves differently across datasets (that is, difference). In each dataset, each group has two important covariates. All nonzero coefficients are equal to 1. The responses are generated from linear regression models with errors standard normally distributed. More details on the simulation setting are provided in the next section. Beyond the proposed CD-SBoost method, we also consider (1) Int-SBoost. This method conducts integrative sparse boosting with multiple datasets (Algorithm 2), but lacks an explicit mechanism to encourage commonality; (2) Sep-SBoost. This method conducts sparse boosting on each dataset separately (Algorithm 1), and then results are combined across datasets. This is a meta-analysis strategy, and there is no consideration of commonality/difference across datasets; and (3) Pool-SBoost. This method combines all datasets directly into one and applies sparse boosting (Algorithm 1). With this approach, all genomic measurements/groups are identified as commonality, and there is no difference identified across datasets. Estimation results for one simulation replicate are shown in Table 1. It is noted that for this specific example, Int-SBoost and Sep-SBoost generate identical estimates. In terms of selection of important covariates, CD-SBoost outperforms Sep-SBoost/Int-SBoost by identifying fewer false positives and outperforms Pool-SBoost by identifying fewer false positives and more true positives as well. The estimation performance is evaluated by RMSE (see the next section for its definition). The RMSE values are 1.217 (CD-SBoost), 1.256 (Sep-SBoost/Int-SBoost), and 3.808 (Pool-SBoost). CD-SBoost has the lowest estimation error. More conclusive simulation results based on multiple replicates are presented in the next section.

**Table 1.** Small example: estimation results of one simulation replicate. Each group of covariates is highlighted in gray.

| | True | | | CD-SBoost | | | Int-SBoost(Sep-SBoost) | | | Pool-SBoost | | |
|---|---|---|---|---|---|---|---|---|---|---|---|---|
| Cov | Data1 | Data2 | Data3 | Data1 | Data2 | Data3 | Data1 | Data2 | Data3 | Data1 | Data2 | Data3 |
| 1 | 1 | 1 | 1 | 0.967 | 0.967 | 0.967 | 0.806 | 1.015 | 1.224 | 1.085 | 1.085 | 1.085 |
| 2 | 1 | 1 | 1 | 1.108 | 1.108 | 1.108 | 1.242 | 1.128 | 0.879 | 1.289 | 1.289 | 1.289 |
| 16 | | | | | | | 0.563 | | | | | |
| | | | | | | | | | | | | |
| 31 | | 1 | | | 0.659 | | | 0.918 | | | | |
| 32 | | 1 | | | 0.497 | | | 0.749 | | | | |
| 33 | 1 | | | 0.686 | | | 0.863 | | | | | |
| 34 | 1 | | | 0.766 | | | 0.974 | | | 0.329 | 0.329 | 0.329 |
| 35 | | | 1 | | | 0.548 | | | 0.568 | 0.42 | 0.42 | 0.42 |
| 36 | | | 1 | | | 0.862 | | | 0.98 | | | |
| 46 | | | | | | | | | | 0.361 | 0.361 | 0.361 |
| 55 | | | | | | | | | | 0.34 | 0.34 | 0.34 |
| | | | | | | | | | | | | |
| 62 | 1 | 1 | | 0.805 | 0.805 | | 1.05 | 0.711 | | 0.698 | 0.698 | 0.698 |
| 63 | 1 | 1 | | 0.878 | 0.878 | | 0.816 | 1.042 | | 0.712 | 0.712 | 0.712 |
| 64 | | | 1 | | | 0.59 | | | 0.591 | | | |
| 65 | | | 1 | | | 0.496 | | | 0.493 | | | |
| 78 | | | | | | | | | | 0.348 | 0.348 | 0.348 |
| | | | | | | | | | | | | |
| 81 | | 1 | 1 | | 1.025 | 1.025 | | 1.114 | 1.296 | 0.779 | 0.779 | 0.779 |
| 82 | | 1 | 1 | | 0.771 | 0.771 | | 0.93 | 0.718 | | | |
| 83 | 1 | | | 0.905 | | | 1.024 | | | 0.475 | 0.475 | 0.475 |
| 84 | 1 | | | 0.768 | | | 0.725 | | | | | |

# 4 Numerical studies

## 4.1 Simulation

We simulate three independent datasets, each with 200 subjects. A total of 1000 covariates with continuous distributions are generated to mimic gene expression data analyzed below. These covariates belong to 20 non-overlapping groups. The sizes of the groups range from 20 to 80 to better mimic practical data. The covariates have marginally normal distributions with means 0 and variances 1. Covariates within the same groups are more strongly correlated than those in different groups. More details on the correlation structure are provided in Appendix B. In each dataset, there are 40 important covariates, with two important covariates per group. Three scenarios for the important covariates/groups are considered: (1) full commonality, where the three datasets share the same set of important covariates and also the same regression coefficients; (2) partial commonality. Here two cases are considered: (a) datasets 1 and 2 have the same set of important covariates and the same regression coefficients, while dataset 3 has a different set of important covariates; and (b) datasets 2 and 3 have the same set of important covariates and the same regression coefficients, while dataset 1 has a different set of important covariates; and (3) no commonality, where none of the important covariates are shared by the three datasets. The 20 groups are randomly allocated into the above three scenarios with proportions $\rho_f$, $\rho_p$, and $\rho_n$.

We consider continuous data under the LR model and censored survival data under the AFT model. With censored survival data, the censoring times are generated independently from uniform distributions. The parameters of censoring distributions are adjusted so that the censoring rates are close to 25%. Under both models, the random errors are generated from $N(0, \sigma^2)$, with $\sigma^2 = 1$ and 3 representing two noise levels. The nonzero regression coefficients are either set all equal to 0.5 or generated randomly. In the latter case, the nonzero regression coefficients in dataset 2 are generated from $\mathcal{U}[0.4, 0.7]$, those specific to dataset 1 are from $\mathcal{U}[0.1, 0.3]$, and those specific to dataset 3 are from $\mathcal{U}[0.8, 1]$. Overall, we have the following simulation settings: (S1) Regression coefficients are randomly generated, and $\sigma^2 = 1$; (S2) Regression coefficients are randomly generated, and $\sigma^2 = 3$; (S3) Regression coefficients are all equal to 0.5, and $\sigma^2 = 1$; and (S4) Regression coefficients are all equal to 0.5, and $\sigma^2 = 3$.

To better gauge performance of CD-SBoost, we compare with Int-SBoost, Sep-SBoost, and Pool-SBoost described above. In addition, two penalization (Lasso) based approaches are also considered: (1) Sep-Lasso, which is similar to Sep-Boost and applies Lasso penalization to each dataset; and (2) Pool-Lasso, which is similar to Pool-SBoost but uses Lasso for estimation.

When evaluating CD-SBoost and alternative approaches, we are the most interested in identification performance. Specifically, we are interested in whether an approach can identify groups behaving the same (i.e. commonality). This is measured using the TP (true positive) and FP (false positive) numbers. More specifically, when computing positive, for a group, we count the number of pairs of distinct datasets where this group behaves the same, and then sum overall groups. For three datasets with full commonality, partial commonality, and difference having proportions $\rho_f$, $\rho_p$ and $\rho_n$, respectively, the true number of positives, denoted as $N_{ig}$, equals $20(2\rho_f + \rho_p)$. CD-SBoost and alternatives conduct variable selection. We also use TP and FP to evaluate variable selection performance. Here the definitions are the same as in the literature. Estimation and prediction performance is also evaluated. Specifically, estimation performance is measured by ERMSE (estimation RMSE), which is defined as $\sqrt{\sum_{m=1}^{M} \| \hat{\boldsymbol{\beta}}^m - \boldsymbol{\beta}^m \|_2^2}$. Prediction performance is measured by PRMSE (prediction RMSE), which is defined as $\sqrt{\sum_{m=1}^{M} \| \hat{\boldsymbol{y}}^m - \boldsymbol{y}^m \|_2^2}$ in an LR model, and as $\frac{1}{\sigma}\sqrt{\sum_{m=1}^{M} (\hat{\boldsymbol{\beta}}^m - \boldsymbol{\beta}^m)^\top \mathrm{cov}(\boldsymbol{X}^m)(\hat{\boldsymbol{\beta}}^m - \boldsymbol{\beta}^m)}$ in an AFT model.

Summary statistics are computed based on 100 replicates. Results for setting S1 and the LR model are presented in Table 2. The rest of the results are presented in Appendix C. Simulation suggests that the proposed CD-SBoost can significantly outperform the alternatives in identifying commonality groups. Specifically, Sep-SBoost and Sep-Lasso can hardly identify commonality. Int-SBoost also does not have a mechanism for identifying commonality. Under Pool-SBoost and Pool-Lasso, all groups are forced to behave the same across datasets. As such, they perform well with the full commonality groups, but fail with the partial commonality and difference groups.

In the identification of important variables, CD-SBoost is observed to have favorable performance. Since Sep-SBoost analyzes each dataset separately, its performance is not strongly affected by the commonality/difference across datasets. Under almost all settings, Sep-SBoost performs worse than CD-SBoost. When the three datasets have a low level of commonality, for example $(\rho_f, \rho_p, \rho_n) = (0.1, 0, 0.9)$, Pool-SBoost has the worst performance, which is as expected. Its performance improves when the level of commonality increases, however, is still inferior to CD-SBoost. Int-SBoost does not show significant advantage over Sep-SBoost in terms of selection. Under all simulation scenarios, the two penalization methods have inferior performance. Specifically, they perform reasonably in terms of true positives but identify too many false positives. In the evaluation of estimation and prediction, the CD-SBoost method is again observed to have favorable performance. The overall observed patterns are similar across simulation settings.

## 4.2 Analysis of leukemia data

Acute myeloid leukemia (AML) is a type of blood and bone marrow cancer. Here data are downloaded from TCGA (https://portal.gdc.cancer.gov/projects/TCGA-LAML). In our analysis, of interest is the regulation of blast count, which is an important diagnostic and prognostic marker, by gene expressions. Data on 173 patients are available for analysis. In the literature, it has been suggested that older patients have disease biology different from that of the young.[23] However, in some of the existing analyses, insufficient attention has been paid to this difference. The patients are divided into two groups: (1) young patients who are younger than 50; and (2) old patients who are 50 years old or older. The sample sizes are 54 and 119, respectively. For this specific example, it is reasonable to expect both commonality (as the two groups have the same cancer) and difference (with the across-age difference suggested in the literature).

In the original data, there are 20,532 gene expressions. In the KEGG pathway database obtained from the Broad Institute, 5266 unique genes represent 186 pathways. In the leukemia dataset, 4243 genes have pathway information and are further considered. This selection is due to consideration on interpretability. The number of leukemia-associated genes is expected to be small to moderate. To improve the reliability of analysis, we further conduct a screening and remove genes whose marginal associations with the response are less significant ($p > 0.01$).

**Table 2.** Simulation under the LR model, scenario S1. In each cell, mean(SD).

| | Variable | | Group | | | |
|---|---|---|---|---|---|---|
| | TP | FP | TP | FP | ERMSE | PRMSE |
| $(\rho_f, \rho_p, \rho_n) = (0.8, 0.2, 0)$, $N_{ig} = 36$ | | | | | | |
| CD-SBoost | **117.4 (0.5)** | **2 (1.4)** | 34.2 (1.3) | 0.1 (0.3) | **0.6 (0.2)** | **23.5 (0.7)** |
| Int-SBoost | 79.8 (6.8) | 40 (8.2) | 0.3 (0.7) | **0 (0)** | 4.1 (0.3) | 39.5 (1.1) |
| Sep-SBoost | 79.6 (6.7) | 40 (8.2) | 0.3 (0.7) | **0 (0)** | 4.2 (0.2) | 39.8 (0.5) |
| Pool-SBoost | 81.1 (6.6) | 31.1(12) | **36 (0)** | 4 (0) | 4.3 (0.3) | 60.8 (3.5) |
| Sep-Lasso | 95.5 (7.9) | 177.1 (34.6) | 0 (0) | 0 (0) | 3.9 (0.3) | 34.4 (5.6) |
| Pool-Lasso | 116.4 (1.1) | 106 (25.9) | **36 (0)** | 4 (0) | 2.7 (0.1) | 37.1 (2.0) |
| $(\rho_f, \rho_p, \rho_n) = (0.6, 0.2, 0.2)$, $N_{ig} = 28$ | | | | | | |
| CD-SBoost | **108.4 (3.2)** | **8.9 (3.1)** | 25 (1.9) | 1 (0.8) | **1.6 (0.3)** | **24.7 (0.7)** |
| Int-SBoost | 81.3 (3.9) | 39.6 (6.7) | 0.1 (0.3) | **0 (0)** | 4.1 (0.2) | 40 (0.8) |
| Sep-SBoost | 81.2 (3.9) | 39.5 (6.7) | 0.1 (0.3) | 0.2 (0.4) | 4.2 (0.2) | 40.6 (1.1) |
| Pool-SBoost | 70.8 (6.7) | 32.4 (8.9) | **28 (0)** | 12 (0) | 4.9 (0.2) | 70.8 (3.8) |
| Sep-Lasso | 97.7 (8.6) | 185 (35.2) | 0 (0) | **0 (0)** | 4 (0.4) | 32.5 (6.1) |
| Pool-Lasso | 102.1 (3.2) | 114.4 (32.2) | **28 (0)** | 12 (0) | 4.2 (0.1) | 52.3 (3.5) |
| $(\rho_f, \rho_p, \rho_n) = (0.1, 0.9, 0)$, $N_{ig} = 22$ | | | | | | |
| CD-SBoost | **102.3 (5.4)** | **17 (7.8)** | 18.9 (1.9) | 0.5 (1.1) | **1.6 (0.6)** | **24.7 (1.5)** |
| Int-SBoost | 78.7 (6.3) | 41.2 (5.5) | 0.1 (0.3) | **0 (0)** | 4.2 (0.1) | 40.5 (0.8) |
| Sep-SBoost | 78.7 (6.3) | 41.2 (5.5) | 0.1 (0.3) | 0.6 (1.1) | 4.3 (0.3) | 41.3 (1.1) |
| Pool-SBoost | 47.3 (3.8) | 43.6 (12.8) | **22 (0)** | 18 (0) | 5.9 (0.2) | 82.5 (2.7) |
| Sep-Lasso | 94.7 (6.6) | 180.8 (29.8) | 0 (0) | **0 (0)** | 3.9 (0.3) | 33.1 (5.3) |
| Pool-Lasso | 75.9 (11.9) | 91.4 (30.6) | **22 (0)** | 18 (0) | 5.6 (0.2) | 71.2 (5.5) |
| $(\rho_f, \rho_p, \rho_n) = (0.4, 0.1, 0.5)$, $N_{ig} = 18$ | | | | | | |
| CD-SBoost | **95.8 (3.9)** | **22.9 (5.4)** | 13.7 (1.7) | 1.9 (1.4) | **2.2 (0.2)** | 26.2 (0.7) |
| Int-SBoost | 79 (7.7) | 43 (8.8) | 0 (0) | **0 (0)** | 4 (0.4) | 39.9 (1.2) |
| Sep-SBoost | 79 (7.7) | 43 (8.8) | 0 (0) | 0.5 (0.7) | 4.1 (0.4) | 39.8 (1.2) |
| Pool-SBoost | 51 (5) | 27.5 (6.4) | **18 (0)** | 22 (0) | 5.5 (0.1) | 79.6 (2) |
| Sep-Lasso | 93.5 (8.9) | 168.1 (63.1) | 0 (0) | **0 (0)** | 3.8 (0.5) | **21.3 (6.1)** |
| Pool-Lasso | 72.5 (7.7) | 78.1 (27.2) | **18 (0)** | 22 (0) | 5.4 (0.2) | 70 (5.1) |
| $(\rho_f, \rho_p, \rho_n) = (0.2, 0.2, 0.6)$, $N_{ig} = 12$ | | | | | | |
| CD-SBoost | **92.5 (5.6)** | 29.2 (8.8) | 8.3 (2.5) | 2.5 (1.2) | **3.3 (0.6)** | **27.2 (1)** |
| Int-SBoost | 77.4 (6.9) | 42.2 (8.1) | 0.1 (0.3) | **0 (0)** | 3.9 (0.4) | 39 (1.3) |
| Sep-SBoost | 77.3 (6.7) | 42.2 (8.1) | 0.1 (0.3) | 0.6 (1) | 4.1 (0.4) | 39 (1.3) |
| Pool-SBoost | 32.8 (4.1) | **26.9 (6.2)** | **12 (0)** | 28 (0) | 6.1 (0.1) | 87.7 (3) |
| Sep-Lasso | 91.1 (8.1) | 170.1 (36.6) | 0 (0) | 0.1 (0.2) | 3.9 (0.5) | 33.9 (7) |
| Pep-Lasso | 53.1 (11.1) | 69.7 (34.7) | **12 (0)** | 28 (0) | 6 (0.2) | 79 (6.2) |
| $(\rho_f, \rho_p, \rho_n) = (0.1, 0, 0.9)$, $N_{ig} = 4$ | | | | | | |
| CD-SBoost | 78.1 (7) | **35 (11.1)** | 2.1 (1) | 3.9 (2) | 4.2 (0.8) | **30.1 (1.8)** |
| Int-SBoost | 76.3 (6.5) | 45.3 (8.4) | 0 (0) | **0 (0)** | 4.4 (0.5) | 38.2 (1.4) |
| Sep-SBoost | 76.3 (6.6) | 45.3 (8.4) | 0 (0) | 0.4 (0.5) | 4.4 (0.5) | 38.4 (1.2) |
| Pool-SBoost | 20.2 (2.8) | **35 (3.9)** | **4 (0)** | 36 (0) | 6.7 (0.6) | 96 (2.9) |
| Sep-Lasso | **88.9 (9.7)** | 159.9 (41.9) | 0 (0) | 0.2 (0.1) | **3.9 (0.6)** | 33.1 (7) |
| Pool-Lasso | 18.6 (6.9) | 58.6 (24.4) | **4 (0)** | 36 (0) | 6.6 (0.1) | 93.3 (6.8) |

Overall, a total of 841 genes are included in analysis. To describe the coordination among genes, we first build a network with edge weights equal to the correlations between two genes. This fully connected network is then filtered into a sparse one using the method in Serrano et al.,[24] which only preserves edges that are statistically significant with respect to a null model for the local assignment of weights to edges. This sparse network is then partitioned into 24 non-overlapping groups using the Louvain method,[25] a fast greedy-optimization-based community detection method.

With CD-SBoost, 22 genes, representing eight groups, are identified in at least one dataset. Out of the identified gene groups, four are identified as commonality. The detailed estimation results are provided in Table 3. A quick literature search suggests that the findings are biologically meaningful. For example, CEP89 encodes protein

**Table 3.** Analysis of leukemia data using CD-SBoost: identified genes and estimates. Genes/groups identified as full commonality are highlighted in gray.

| Gene | Data 1 | Data 2 | Gene | Data 1 | Data 2 |
|---|---|---|---|---|---|
| CEP89 | 0.136 | 0.136 | C6ORF89 | −0.084 | −0.084 |
| DPYSL3 | 0.044 | 0.044 | | | |
| | | | APH1B | −0.185 | |
| ATXN1 | −0.067 | −0.123 | DIP2B | | −0.227 |
| CLTCL1 | −0.150 | −0.091 | C3ORF49 | | 0.229 |
| CMA1 | −0.223 | −0.052 | LOC158267 | −0.167 | |
| CSF3R | | −0.174 | HSA6077 | | 0.095 |
| ACTN2 | | −0.135 | | | |
| | | | ADA | 0.081 | 0.081 |
| CD27 | −0.065 | −0.175 | | | |
| COL4A3 | | −0.116 | CYB5R1 | −0.102 | −0.102 |
| DSC1 | −0.091 | −0.065 | | | |
| CDHR3 | | −0.106 | C14ORF123 | −0.247 | |
| CDC25B | | −0.137 | ALDH9A1 | −0.146 | −0.090 |

**Table 4.** Analysis of leukemia data: numbers of overlapping genes.

| | CD-SBoost | Int-SBoost | Sep-SBoost | Pool-SBoost | Sep-Lasso | Pool-Lasso |
|---|---|---|---|---|---|---|
| CD-SBoost | 22 | 7 | 7 | 3 | 14 | 1 |
| Int-SBoost | | 17 | 17 | 2 | 12 | 1 |
| Sep-SBoost | | | 17 | 2 | 12 | 1 |
| Pool-SBoost | | | | 4 | 2 | 1 |
| Sep-Lasso | | | | | 169 | 2 |
| Pool-Lasso | | | | | | 3 |

Centrosomal Protein 89, which is found to be involved in mitochondrial metabolism and neuronal functions.[26] Both CD27 and CDC25B have been found to play an important role in AML. CD27 protein is a member of the TNF-receptor superfamily, which regulates T cell generation and immunity. Studies have shown that CD27 is elevated in the sera of AML patients and associated with poor survival.[27] CDC25B is a member of the CDC25 family of phosphatases. It is involved in cell cycle progression, chromatin modulation, and transcriptional regulation.[28] CDC25 is a candidate prognostic marker and therapeutic target not only for AML but also for other types of cancers.[29] Adenosine deaminase 1 (ADA) has been found to be a diagnostic and prognostic marker for patients with acute lymphoblastic leukemia.[30] CSF3R is identified only in the older patients. It encodes a protein which is the receptor for colony stimulating factor 3 and works as a key regulator of granulocytes.[31] Abnormal expressions of CSF3R have been found in severe congenital neutropenia, chronic neutrophilic leukemia, and other types of cancer.[31]

The data is also analyzed using the alternatives described above. The summary comparison results are provided in Tables 4 and 5. Overall, it is found that different approaches lead to different findings. Sep-SBoost and Int-SBoost identify the same set of 17 genes in eight groups. Sep-Lasso identifies 169 genes in 23 groups. With these three methods, all groups (with nonzero effects) are identified as behaving differently across the two datasets. Pool-SBoost identifies four genes from four groups, and Pool-Lasso identifies three genes from three groups. Detailed estimation results using the boosting alternatives are provided in Appendix C, and those using the Lasso alternatives are provided in *Supplementary materials*.

In practical data analysis, it is challenging to objectively determine which set of results is more sensible. To complement the identification/estimation analysis, we evaluate prediction and stability, which may provide support to the analysis results to a large extent: if the identification/estimation results are reliable, prediction is expected to be more accurate and stability to be higher. Specifically, each dataset is randomly split into a training and a testing set with sizes 3:1. Estimation is conducted using the training set and used to make prediction for subjects in the testing set. The prediction RMSE is then computed to evaluate prediction performance.

**Table 5.** Analysis of leukemia data: numbers of identified gene groups, commonalities and differences.

| | Groups | Commonalities | Differences |
|---|---|---|---|
| CD-SBoost | 8 | 4 | 4 |
| Int-SBoost | 8 | 0 | 8 |
| Sep-SBoost | 8 | 0 | 8 |
| Pool-SBoost | 4 | 4 | 0 |
| Sep-Lasso | 23 | 0 | 23 |
| Pool-Lasso | 3 | 3 | 0 |

**Table 6.** Analysis of kidney cancer data using CD-SBoost: identified genes and estimates.

| Gene | Data 1 | Data 2 | Data 3 | Gene | Data 1 | Data 2 | Data 3 |
|---|---|---|---|---|---|---|---|
| OPH | 0.129 | 0.129 | | HYAL2 | | | 0.093 |
| ASRGL1 | 0.109 | 0.109 | | IDH2 | 0.046 | 0.046 | |
| ARHGEF6 | | | 0.063 | | | | |
| BDNFAS | | | 0.202 | KCNF1 | −0.131 | −0.131 | −0.131 |
| RUNX1IT1 | | | −0.741 | LOC154339 | −0.069 | −0.069 | −0.069 |
| CEP170 | 0.258 | | | NMB | −0.136 | −0.136 | −0.136 |
| CEP76 | | | −0.003 | | | | |
| CIB1 | 0.053 | | | LOC100422366 | −0.091 | −0.091 | |
| CIRBP | | | −0.003 | PIP5K1A | | | −0.087 |
| CLASP1 | | | 0.556 | PRR5ARHGAP8 | | | 0.249 |
| CNNM2 | | −0.105 | | SLC25A26 | | | 0.059 |
| CSNK2A1 | | | −0.052 | SMC3 | | | −0.072 |
| ACHM1 | −0.112 | | | SPNS2 | 0.144 | 0.047 | |
| FLJ21736 | | 0.078 | | SUCLG2 | | | 0.155 |
| HSPC139 | | | −0.138 | TAC2 | | 0.180 | |
| LOC101928965 | 0.102 | | | TARS2 | | | −0.140 |
| | | | | LOC651293 | −0.224 | | |
| EBF1 | | | 0.091 | MIEF1 | | 0.200 | |
| EFCAB3 | 0.168 | 0.168 | 0.168 | | | | |
| | | | | TMPRSS9 | −0.101 | −0.101 | −0.101 |
| FAIM2 | | | 0.192 | | | | |
| FIS1 | | −0.279 | | TXLNG | −0.162 | −0.162 | −0.162 |
| FKBP1A | | | −0.187 | | | | |
| FLJ16103 | | 0.168 | | VPS33B | 0.173 | | |
| FLJ44635 | | | 0.216 | XRN2 | −0.064 | −0.162 | −0.162 |
| FUT3 | | −0.434 | | YAF2 | | −0.188 | −0.188 |
| FUT6 | 0.177 | | | | | | |
| GALC | | −0.184 | | ZNF121 | | | −0.132 |
| HES4 | | −0.190 | | | | | |
| HIGD1B | | | 0.176 | | | | |
| HIST1H2BM | | −0.147 | | | | | |

Note: Genes/groups identified as full commonality are highlighted in gray. Those identified as partial commonality are highlighted in light gray.

This process is repeated 100 times. The average RMSE values are 6.159 (CD-SBoost), 6.551 (Int-SBoost), 6.620 (Sep-SBoost), 6.859 (Pool-SBoost), 9.605 (Sep-SLasso), and 6.418 (Pool-SLasso), respectively. With this random splitting approach, we also evaluate the stability of findings. Specifically, the Observed Occurrence Index (OOI) values,[a] which are the probabilities of a finding being repeated, are 0.788 (CD-SBoost), 0.762 (Int-SBoost), 0.760 (Sep-SBoost), 0.264 (Pool-SBoost), 0.721 (Sep-SLasso) and 0.173 (Pool-SLasso), respectively. Note that the overlapping genes identified by Sep-Lasso and CD-SBoost usually have rather high OOIs, which leads to not too much difference in their OOIs. Overall, the improved prediction and stability support the validity of our analysis.

**Table 7.** Analysis of kidney cancer data: numbers of overlapping genes.

| | CD-SBoost | Int-SBoost | Sep-SBoost | Pool-SBoost | Sep-Lasso | Pool-Lasso |
|---|---|---|---|---|---|---|
| CD-SBoost | 51 | 24 | 24 | 11 | 35 | 19 |
| Int-SBoost | | 49 | 49 | 14 | 38 | 19 |
| Sep-SBoost | | | 49 | 14 | 38 | 19 |
| Pool-SBoost | | | | 23 | 16 | 12 |
| Sep-Lasso | | | | | 238 | 47 |
| Pool-Lasso | | | | | | 95 |

**Table 8.** Analysis of kidney cancer data: numbers of identified gene groups, commonalities and differences.

| | Groups | Commonalities | Differences |
|---|---|---|---|
| CD-SBoost | 12 | 4 | 8 |
| Int-SBoost | 16 | 0 | 16 |
| Sep-SBoost | 16 | 0 | 16 |
| Pool-SBoost | 16 | 16 | 0 |
| Sep-Lasso | 30 | 0 | 30 |

## 4.3 Analysis of kidney cancer data

TCGA has three kidney cancer datasets, on kidney renal clear cell carcinoma (KIRC), kidney renal papillary cell carcinoma (KIRP), and chromophobe kidney cancer (KICH), respectively. In this analysis, the outcome of interest is overall survival. For each dataset, we exclude the subjects with neo-adjuvant therapy, missing in survival outcome and gene expressions. The final sample sizes are 65 with nine deaths (KICH), 291 with 44 deaths (KIRP), and 517 with 165 deaths (KIRC), respectively. As all three datasets are on kidney cancer, commonality is expected. With the (high) heterogeneity of cancer, difference is also expected. Data processing is conducted in a similar manner as described above, which screens 2192 genes out of 20,534. With the same approach as described above, these genes are divided into 40 non-overlapping groups.

CD-SBoost identifies 51 genes representing 12 groups. Among those groups, four are identified as full commonality, four are identified as partial commonality, and the rest are identified as difference. Detailed estimation results are provided in Table 6. It is again found that the identification results are meaningful. For example, gene NMB is identified for all three subtypes. It encodes protein Neuromedin B and has been found to be a growth factor in cancer cells.[32] Inhibition of the Neuromedin B signaling is a possible new target in cancer treatment.[33] VPS33B is identified in KIRC and a known tumor suppressor in liver cancer. Gene GALC is identified in KIRP and a prognostic marker associated with both tumor and distant metastasis.[34] Gene CSNK2A1, which is only identified in KICH, is a cancer prognostic marker.[35]

Data is also analyzed using the alternatives. Tables 7 and 8 again suggest that different approaches lead to different findings. Detailed estimation results using the boosting based approaches are provided in Appendix C, and those using the Lasso-based approaches are provided in *Supplementary materials*. In prediction evaluation, as the outcome variable is censored survival, the logrank statistic is used. The average logrank statistics are 10.322 (CD-SBoost), 9.287 (Int-SBoost), 8.987 (Sep-SBoost), 8.823 (Pool-SBoost), 9.453 (Sep-SLasso), and 8.077 (Pool-SLasso), respectively. In stability evaluation, the OOI values are 0.907 (CD-SBoost), 0.854 (Int-SBoost), 0.854 (Sep-SBoost), 0.635 (Pool-SBoost), 0.882 (Pool-Lasso), and 0.912 (Sep-Lasso). Overall, the proposed CD-SBoost is found to have competitive performance.

# 5 Conclusion

With the fast accumulation of cancer omics data, there is a growing popularity of across-cancer analysis. This study has explicitly focused on the commonality and difference across cancer patient groups, which has been considered in very limited literature. Complementing the existing penalization and other studies, CD-SBoost, a novel sparse boosting approach, is developed. Under simpler settings, it has been shown that boosting, penalization, and other techniques have specific advantages and do not dominate each other. Considering the satisfactory properties of boosting under other settings, the proposed methodological research is worthwhile. Comparing with the existing approach, the proposed CD-SBoost introduces a new penalty to promote commonality and, meanwhile, allow for difference. The newly proposed penalty has some similarity with the one in Huang et al.[15] However, there are key differences. Specifically, the penalty in Huang et al. is concerned with sparsity structure and cannot promote equal regression coefficients. A novel computation of increments is proposed, which determines all increments simultaneously. In addition, the proposed analysis is conducted at the group level. That is, a whole group can be identified as either commonality or difference. This has been designed to accommodate the coordination among genomic measurements. These innovations in boosting technique may also benefit studies.

Under a wide spectrum of simulation settings, CD-SBoost is observed to have favorable performance. In data analysis, it has found significant differences across cancer patient groups, which may suggest the limitation of the existing analyses that assume homogeneity. Overall, this study may provide a practically useful new venue for analyzing cancer omics data under heterogeneity.

Beyond those considered in this article, there are many other types of cancer outcomes/models. It will be of interest to extend the proposed analysis to other outcomes/models. In the proposed analysis, the patient groups are assumed to be pre-defined. It will be of interest to examine whether it is possible to couple the proposed approach with data-dependent patient group identification. Data analysis results may also worth further investigation.

## Declaration of conflicting interests

The author(s) declared no potential conflicts of interest with respect to the research, authorship, and/or publication of this article.

## Funding

The author(s) disclosed receipt of the following financial support for the research, authorship and/or publication of this article: This study was supported by National Natural Science Foundation of China (11605288, 71771211), Fund for building world-class universities (disciplines) of Renmin University of China, and National Institute of Health (CA204120, CA216017).

## Note

a. We, firstly, compute OOI of each gene, and average over the 15 highest OOIs.

## ORCID iD

Yifan Sun https://orcid.org/0000-0001-7437-6244
Yang Li https://orcid.org/0000-0002-6287-5094
Shuangge Ma https://orcid.org/0000-0001-9001-4999